\documentclass{aa}

\usepackage{graphicx}
\usepackage{txfonts}
\usepackage[dvipsnames]{xcolor}
\usepackage{xparse,xcolor}
\begin{document}

   \title{Hystereses in dwarf nova outbursts and low-mass X-ray binaries}

   \author{J.-M. Hameury
          \inst{1}
          \and
          J.-P. Lasota\inst{2,3}
          \and
          C. Knigge\inst{4}
          \and
          E.G. Körding\inst{5}
          }

   \institute{Université de Strasbourg, CNRS, Observatoire Astronomique de Strasbourg, UMR 7550, 67000 Strasbourg, France\\
             \email{jean-marie.hameury@astro.unistra.fr}
         \and
         Institut d'Astrophysique de Paris, CNRS et Sorbonne Universit\'es, UPMC Paris~06, UMR 7095, 98bis Bd Arago, 75014 Paris, France
\and
             Nicolaus Copernicus Astronomical Center, Polish Academy of Sciences, Bartycka 18, 00-716 Warsaw, Poland
              \and
              School of Physics and Astronomy, University of Southampton, Southampton SO17 1BJ, UK
              \and
              Department of Astrophysics/IMAPP, Radboud University, PO Box 9010, NL-6500 GL Nijmegen, the Netherlands           
}

   \date{}


  \abstract
   {The disc instability model (DIM) successfully explains why many accreting compact binary systems exhibit outbursts, during which their luminosity increases by orders of magnitude. The DIM correctly predicts which systems should be transient and works regardless of whether the accretor is a black hole, a neutron star or a white dwarf. However, it has been known for some time that the outbursts of X-ray binaries (which contain neutron-star or black-hole accretors) exhibit hysteresis in the X-ray hardness-intensity diagram (HID). More recently, it has been shown that the outbursts of accreting white dwarfs also show hysteresis, but in a diagram combining optical, EUV and X-ray fluxes.}
   {We examine here the nature of the hysteresis observed in cataclysmic variables and low-mass X-ray binaries.}
   {We use the Hameury et al. (1998) code for modelling dwarf nova outbursts, and construct the hardness intensity diagram as predicted by the disc instability model.}
   {We show explicitly that the standard DIM -- modified only to account for disc truncation -- can explain the hysteresis observed in accreting white dwarfs, but cannot explain that observed in X-ray binaries.}
   {The spectral evidence for the existence of different accretion regimes / components (disc, corona, jets, etc.) should be based only on wavebands that are specific to the innermost parts of the discs, i.e. EUV and X-rays, which is a difficult task because of interstellar absorption. The existing data, however, indicate that an EUV/X-ray hysteresis is present in SS Cyg.}


   \maketitle
%

\section{Introduction}

Dwarf novae (DNs) are a subclass of cataclysmic variables (CVs) which undergo recurrent outbursts usually lasting a few days. Outbursts are separated by quiescent intervals of a few weeks \citep[see e.g.][for an encyclopaedic review of CVs]{w95}. It is now widely accepted that these outbursts are the result of a thermal-viscous instability of the accretion disc which arises when hydrogen becomes partially ionized and the opacities depend sensitively on temperature \citep[see][for a review of the model]{l01}. A stability analysis shows that when the local mass transfer rate $\dot{M}$ is in the range $[\dot{M}_{\rm crit}^- , \dot{M}^+_{\rm crit}]$ the disc is thermally and viscously unstable; for lower values of $\dot{M}$, the disc in stable on a cold branch, and for higher values of $\dot{M}$, the disc is also stable, but on a hot branch.

In quiescence, mass accumulates in the disc which slowly heats up, the local mass transfer rate increases until it reaches at some point in the disc the critical $\dot{M}_{\rm crit}^-$ for which the instability develops. The instability propagates throughout the disc via two heating fronts moving inwards and outwards; the outer heating front eventually reaches the outer edge of the disc which is then almost in a steady state with a mass accretion rate onto the white dwarf larger than the mass transfer rate from the secondary. The disc empties until a cooling front starts propagating at the outer edge of the disc an brings the system back to quiescence.

Confronting the predictions of the disc instability model (DIM) with the observations has been a long standing endeavour, with of course the objective of validating the model, but also of putting constraints on the viscosity which still remains poorly understood, despite spectacular progresses with the realization that angular momentum transport is most probably the result of the magnetorotational instability \citep{bh91}. In the DIM, viscosity is still modelled according to the \citet{ss73} $\alpha$ parametrization. Whereas $\alpha$ is reasonably well constrained when the disc is hot, there are still large uncertainties when the disc is in quiescence. A new DIM version using viscosity and disc vertical structures obtained through MRI simulation has been recently developed by \citet{Colemanetal16} but also in this case the cold disc physics is causing problems.

Colour variations should be one of the most obvious observational features that models should aim at reproducing. It has been known for long that dwarf novae follow a loop in the $U-B$, $B-V$  colour plane \citep[e.g.][for the cases of \object{SS Cyg} and \object{VW Hyi}]{b80}. This fact seems to have been somehow forgotten, perhaps because the precise modelling of the optical colours of the accretion disc is not easy, as it requires in principle calculations of detailed spectra, and also because it could be accounted for as well by the DIM and by the mass transfer instability model \citep{bp81}. A similar phenomenon is also seen when comparing the visible and EUV/FUV light curves, and is usually described in terms of the so-called UV delay: the UV rise is delayed by approximately 0.5 day in a system such as \object{SS Cyg} \citep{cwp86}; it was initially thought that the importance of the delay was a good indicator of the place where the instability initially develops \citep{s98}: the so-called inside-out outbursts for which the instability develops close to the inner disc edge should exhibit shorter UV delays than the outside-in outbursts for which the instability sets in further away. \citet{shl03} showed that the situation is in fact more complex, that disc truncation plays a major role and that, contrary to expectations, the delay for outside-in outbursts is usually shorter than for inside-out outbursts.

\begin{figure}
   \centering
   \includegraphics[width=\columnwidth]{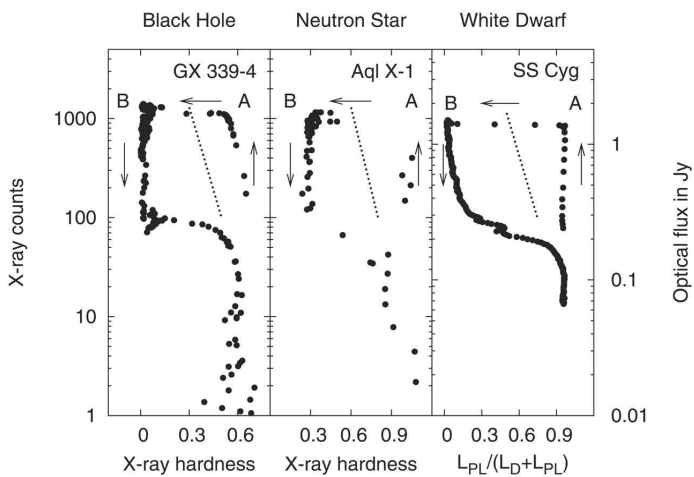}
   \caption{Hardness-intensity diagrams for a black hole, a neutron star and a cataclysmic variables. Here, $L_{\rm PL}$ and $L_{\rm D}$ are is the X-ray and EUV luminosities respectively. Data is taken from \citet{mb04} and \citet{wmm03}. Figure reprinted from \citet{krkf08} by permission of {\it Science}.}
   \label{fig:science_paper}%
   \end{figure}

\citet{krkf08} showed the temporal evolution of the optical flux vs. the EUV/X-ray ratio in SS Cyg: in quiescence, most of the flux is emitted in the X-ray band; during an outburst, the optical flux rises first; close to the outburst maximum, the X-ray flux vanishes while the EUV flux reaches its maximum, and during decline the optical flux decreases whereas the relative contribution of X-rays increases up to its quiescent  value. While this was not discussed in the paper, the data shows a hysteresis in the optical/hardness plot, very similar to the one observed in LMXBs; Fig. \ref{fig:science_paper} reproduces their Fig. 1. \citet{krkf08} argued that the EUV/X-ray ratio is a measure of the optical depth of the inner parts of the accretion flow. They suggested, that similar to XRBs CVs should emit the most prominent radio emission at the initial rise and during the transition from optically thin to optically thick inner accretion flow (in CVs, it is presumably the boundary lager that changes optical depth; in LMXBs, the transition optically thin / optically thick would rather -- certainly for LMXBs containing black holes -- takes place in the disc). This is the phase where the authors detected the source in the radio bands.

The interpretation of the similarity between the HIDs for LMXBs and CVs is, however, not as straightforward as it might seem. Low-mass X-ray binaries (LMXBs) are also subject to the same instability as dwarf novae \citep{kr98,dhl01} which produces soft X-ray transients (SXTs) \citep[see e.g.][for review of the observations]{ts96,csl97,yy15,tsh16}. The differences between SXTs and DNs are due to the nature of the compact object, which results in strong irradiation of the accretion disc during SXT outbursts, together with larger accretion discs de to both a larger companion mass and longer orbital periods in LMXBs. These account for the much longer recurrence time between outbursts (years or decades instead of months) and longer outburst durations (months instead of a few days). The compactness of the accreting object is also responsible for most of the emission in SXTs to peak in the X-ray domain instead of the UV/EUV. In LMXBs, hard and soft X-rays originate from the innermost part of the accretion flow, whereas in cataclysmic variables, X-rays are emitted at the very vicinity of the white dwarf, but the optical light is emitted by the whole disc. The observed hysteresis therefore relates emission from the same region (or from very nearby regions) in LMXBs, whereas this is not at all the case in CVs.

In this paper, we reassess the multi-wavelength time evolution of dwarf novae as predicted by the disc instability model, with a particular emphasis on SS Cyg, for which we have the most comprehensive set of observations, and for which the parameters (distance, orbital parameters) are reasonably well determined, and compare our results with observations. We also provide graphs showing where, according to the DIM, the majority of the emitted light originates, as a function both of time and wavelength. We show that the DIM naturally accounts for the observed hysteresis when correlating X-ray and optical emission, whereas it cannot explain the hysteresis observed in LMXBs when plotting the X-ray flux versus the X-ray hardness. We show, however, that the existing data for SS Cyg suggest that a hysteresis could also exist when considering the X-ray plus EUV flow, presumably representative of the accretion rate in the inner disc, versus the X-ray / EUV hardness ratio. This hysteresis, if confirmed, cannot be explained by the DIM alone; it would be similar to the one observed in LMXBs and could point to similar mechanisms.

\section{The model}

We use here the version of the DIM as described in \citet{hmdlh98} in which heating by the tidal torque and stream impact have been incorporated \citep{bhl01}. We also take into account irradiation of the accretion disc by the compact object, which impacts the outburst cycle by modifying  the vertical disc structure and thus the S-curve, as well as the emitted spectra because of reprocessing. For the spectral modelling, we follow the procedure described in \citet{shl03}, with the difference that the disc spectrum is obtained by summing blackbodies instead of \citet{k79} spectra. Given the small differences found in \citet{shl03} between these two different approximations, and given the fact that the spectrum of an accretion disc is not very well represented locally by a Kurucz spectrum, this is a simple and fair assumption. A detailed calculation as done in \citet{ilhs10} for time-dependent accretion discs would be CPU intensive and clearly outside of the scope of this paper.

In addition to the flux emitted by the accretion disc, we take into account the contribution of the boundary layer, which is assumed to emit hard X-ray when optically thin; we assume that this happens when the accretion rate is below $2 \; 10^{16}$ g s$^{-1}$. This value is twice as large as the one used in \citet{shl03} to account for the smaller white dwarf mass and hence larger emitting surface we assume here (see below). In order to obtain a smooth transition, we assume that the X-ray flux is cut by an exponential factor $\exp(-\dot{M}_{\rm acc}/2 \; 10^{16}\rm g\;s^{-1})$. When optically thick, the spectrum is that of a blackbody, and we assume that the emitting area is $f_{\rm em} 4 \pi R_{\rm wd}^2$, with $f_{\rm em}$ given by \citep{pr85,shl03}:
\begin{equation}
f_{\rm em} = 10^{-3} \left( \frac{\dot{M}_{\rm acc}}{10^{16} \rm gs^{-1}} \right)^{0.28}
\end{equation}
where $R_{\rm wd}$ is the white dwarf radius, and $\dot{M}_{\rm acc}$ is the accretion rate onto the white dwarf. 

We also include the contribution of the white dwarf, assumed to be a blackbody with a constant temperature, and that of the fraction of the hot spot luminosity which is not included in the disc model; we assume that the hot spot radiates as a blackbody with a temperature of 10,000 K. We assume that one half of the stream impact energy is emitted by the hot spot. We finally include the contribution from the secondary star, which is the sum of two blackbodies with different effective temperatures but the same emitting area: the unilluminated backside of the secondary, plus the contribution of the illuminated side of the secondary which is assumed to have a temperature $T_2$ given by:
\begin{equation}
\label{eq:t2}
T_2^4 = T_*^4  + \left( \frac{R_{\rm wd}}{a} \right)^2 \left(T_{\rm BL}^4 f_{\rm em} + T_{\rm wd}^4 \right)
\end{equation}
where $T_*$, $T_{\rm BL}$ and $T_{\rm wd}$ are the secondary, boundary layer and white dwarf temperatures respectively, and $a$ is the orbital separation. Irradiation by the disc is not included in Eq. (\ref{eq:t2}) since the disc luminosity is emitted perpendicular to the disc plane and only a small fraction of it can contribute to the heating of the secondary. The unilluminated temperature is taken to be 4000 K. As for the disc, one could have assumed the secondary spectrum to be given by \citet{k79}; again, given the approximations made here, in particular the fact that the temperature of the irradiated side of the secondary is constant over the whole hemisphere instead of decreasing from the equator to the pole, this refinement is not appropriate, and we therefore use a blackbody spectrum.

\begin{table}
\caption{Binary parameters for SS Cyg} 
\label{tab:sscyg}
\centering 
\begin{tabular}{l c c c} 
\hline\hline 
Parameter & Value & References \\ 
\hline 
$P_{\rm orb}$ (hr) & 6.603 & 1 \\
$M_{\rm wd}$ (M$_\odot$) & 1.00 & 2 (see text) \\
$M_2/M_{\rm wd}$  & 0.67 & 2 \\
$R_{\rm tid}$ (cm) & $5.17 \times 10^{10}$ cm \\
$R_{\rm wd}$ (cm) & $5.5 \times 10^{8}$ cm \\
Distance (pc) & 114 & 3 \\
\hline
\end{tabular}
\tablebib{
(1)~\citet{rk03}; (2)~\citet{brb07}; (3) \citet{mj13}
}
\end{table}

We use the orbital parameters given in Table \ref{tab:sscyg}. We have chosen $M_1 =1$ M$_\odot$, which does not correspond to the fiducial value given by \citet{brb07} of 0.81 M$_\odot$ and is significantly smaller than the previously used value of 1.2 M$_\odot$ because with a primary mass of 0.81 M$_\odot$, the disc extension is small and the peak luminosity cannot reach the observed value. $M_1 = 1.0$ M$_\odot$ corresponds to the upper range of allowed values $[0.62 - 1.00]$ M$_\odot$ provided by \citet{brb07}. The value of the radio parallax determined by \citet{mj13} also requires that the primary mass to be close to the mass of the Sun.

\section{Results for accreting white dwarfs}

   \begin{figure}
   \centering
   \includegraphics[angle=-90,width=\columnwidth]{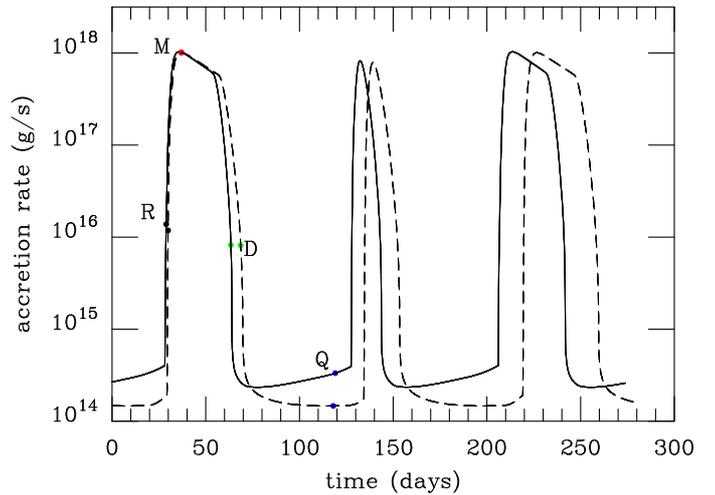}
   \caption{Light curve obtained using parameters given in Table 1. The solid line corresponds to the unilluminated case; the dashed line to the maximally irradiated case. Snapshots in Figs. \ref{fig:spectra} -- \ref{fig:optical} are taken at phases labelled R, M, D, Q.}
   \label{fig:lightcurve}%
   \end{figure}
   
   \begin{figure}
   \centering
   \includegraphics[angle=-90,width=\columnwidth]{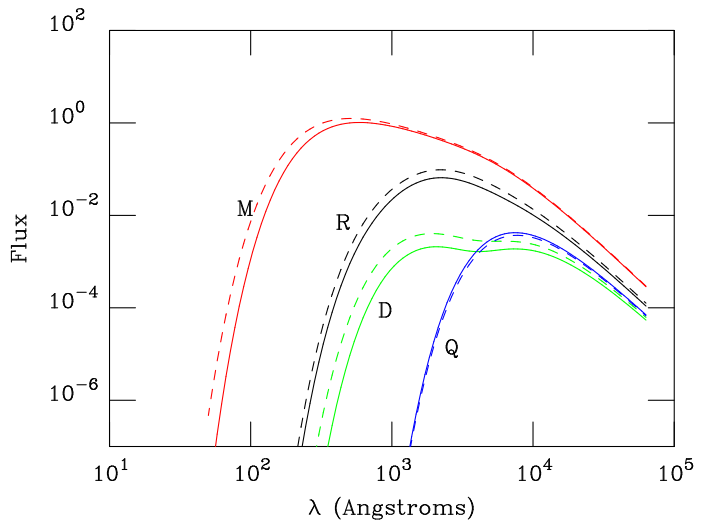}
   \caption{Energy spectra emitted by the accretion disc at times indicated in Fig. \ref{fig:lightcurve}. The black curve corresponds to the rise to maximum (R), the red one to the outburst maximum (M), the green one to decay (D), and the blue one to quiescence (Q). The solid line corresponds to the unilluminated case; the dashed line to the maximally irradiated case. Note that contributions from other components (white dwarf surface, boundary layer, hot spot and secondary star have not been included here. }
   \label{fig:spectra}%
   \end{figure}
     
   \begin{figure}
   \centering
   \includegraphics[angle=-90,width=\columnwidth]{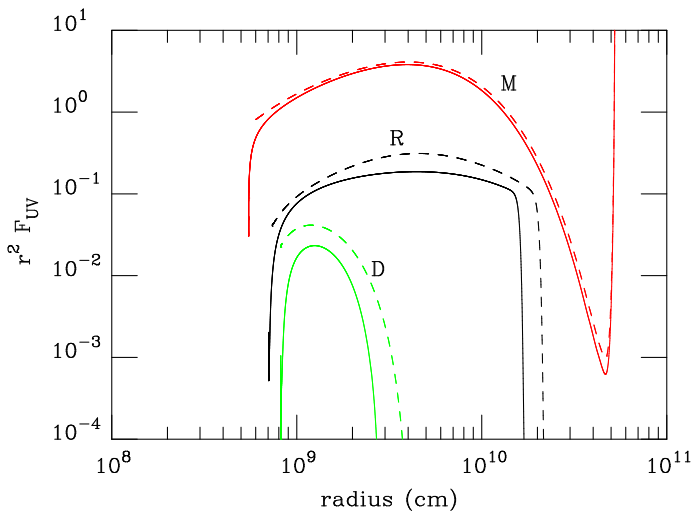}
   \caption{Radial distribution of the UV flux during rise, maximum and decay. The colour coding is the same as in Fig. \ref{fig:optical}. The solid line corresponds to the unilluminated case; the dashed line to the maximally irradiated case. The UV flux in quiescence is vanishingly small and is not shown here.}
   \label{fig:uv}%
   \end{figure}
   
   \begin{figure}
   \centering
   \includegraphics[angle=-90,width=\columnwidth]{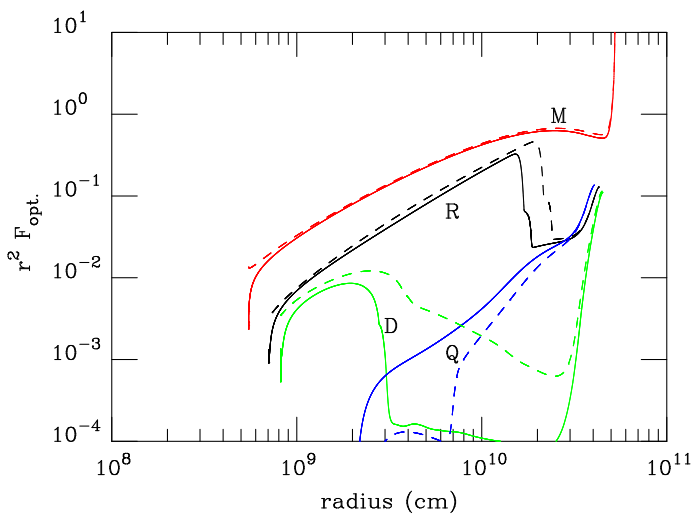}
   \caption{Radial distribution of the optical flux at times shown in Fig. \ref{fig:lightcurve}. The black curve corresponds to the rise to maximum (R), the red one to the outburst maximum (M), the green one to decay (D), and the blue one to quiescence (Q). The solid line corresponds to the unilluminated case; the dashed line to the maximally irradiated case.}
   \label{fig:optical}%
   \end{figure}
   
We first consider the case where the disc is truncated by e.g. a magnetic field and does not extend all the way down to the white dwarf surface during quiescence, which provides the best fit for the UV delay in SS Cyg \citep{shl03}. We take the truncation radius to be:
\begin{equation}
  r_{\rm in} = 5.21 \times 10^8  \mu_{30}^{4/7} M_1^{-1/7} \left( \frac{\dot{M}_{\rm acc}}{10^{16} \; \rm g s^{-1}} \right)^{-2/7} \; \rm cm,
\end{equation} 
which corresponds to the case where truncation is caused by a dipolar magnetic field with moment $\mu_{30}$ in units of $10^{30}$ G cm$^3$, and $M_1$ is the primary mass in solar units. 

We have considered two extreme cases for the irradiation of the disc: the case where irradiation is neglected, and the case where it is treated as in \citet{hlw00}, i.e. the irradiation flux impinging onto the disc is given by:
\begin{equation}
\sigma T_{\rm ill}^4 = \frac{GM_{\rm wd}\dot{M_{\rm acc}}}{8 \pi R_{\rm wd}}^3 {1 \over \pi} [\arcsin \rho -\rho (1-\rho^2)^{1/2} ]
\label{eq:ill}
\end{equation}
where $\rho = R_{wd}/r$, and $r$ is the radial coordinate in the disc. This assumes that the energy released by accretion onto the white dwarf is emitted isotropically by the white dwarf surface. It is a fair approximation when the disc is truncated. It clearly is an overestimate when an optically thick boundary layer forms, since only a fraction of this energy will be thermalised by the whole white dwarf surface. Note that irradiation by a hot white dwarf decreases as $r^{-3}$; it is quite significant in the inner parts of the disc because the white dwarf extends very much above the disc surface, but the outer parts of the disc are not much affected by irradiation. This contrasts with the LMXB case, for which this effect does not exist. In LMXBs, irradiation is observed to be responsible for most of the optical light during outbursts or in steady systems. In this case irradiation can vary as $r^{-2}$, but with a relatively small coefficient \citep[][assume that $\sigma T_{\rm ill}^4 = \mathcal{C} L_{\rm x}/4 \pi r^2$ with $\mathcal{C} \sim 5 \times 10^{-3}$]{dlh99}. As viscous heating varies as $r^{-3}$, irradiation will always overcome viscous heating at large distance if the disc is large enough. This is the case in LMXBs; in CVs, using the same value for $\mathcal{C}$ does not lead to significant effects, because of the smaller disc extent and of the smaller accretion efficiency. Our numerical simulations have shown that this is indeed the case.

The resulting light curve is presented in Fig. \ref{fig:lightcurve}. It shows an alternation of long and short outbursts, as observed in SS Cyg. The irradiated and unirradiated cases are very similar; when irradiation is taken into account, the major outbursts last slightly longer, the quiescence level is reduced because the disc mass is smaller at the end of an outburst, and the recurrence time is slightly increased. We show in Fig. \ref{fig:spectra} the disc spectra obtained at 4 different epochs, representative of the rise to maximum, maximum, decay and quiescence. Contributions from other components of the system (white dwarf, secondary, boundary layer, hot spot) have not been included here; except for the boundary layer, their contribution is significant only during quiescence. Irradiation has little effect on these spectra; the main differences are that they extend to slightly lower wavelengths and the fluxes are slightly higher in the irradiated case, despite the accretion rate being almost identical, except in quiescence. Figs. \ref{fig:uv} and \ref{fig:optical} show the contribution of each ring in the disc to the UV (1250 \AA) and optical (5500 \AA) fluxes.  The position of the maximum in $r^2 F_\nu$ indicates the region of the disc which is the main contributor to the flux, provided that $F_\nu (r)$ varies on scales of order of $r$, which is obviously not the case for the outermost regions of the disc, where heating is dominated by dissipation of tidal forces. Despite the large value of $r^2 F_\nu$ at the outer edge, these regions contribute less than 1\% to the total UV flux throughout the whole cycle, and to 10\% of the optical flux at maximum but 60\% during decay. As can be seen, in the hot state, the UV-domain  $r^2F_\nu$ curve is rather flat over most of the disc, and all regions contribute more or less evenly to the UV flux. This contrasts with the optical wavelengths where most of the flux originates from the outermost ring in the hot state. It can also be seen that differences between the irradiated and non-irradiated cases exist in optical during decay. They appear in regions of the disc which have returned to the cold state, and do not contribute much to the optical light anyway. Differences are also in the contributions of the inner disc to the optical light during quiescence, which are also small; these are consistent with the reduced $\dot{M}_{\rm acc}$ in the irradiated case.

These results could in principle be compared with observations of dwarf novae using eclipse mapping techniques \citep[see e.g.][and references therein]{b16}. The examination of Fig. 3 in \citet{bc01} showing the radial distribution of the intensity of the dwarf nova \object{EX Dra} shows, however, a number of differences with the picture presented here. In particular, in quiescence, most of the optical light is emitted by the innermost part of the accretion disc, with a significant, though smaller, contribution from an outer ring which is not axisymmetric. One should keep in mind that, because this technique is limited to high inclination systems, some artefacts may be present \citep{s94}. Irradiation of the inner disc by the white dwarf can also be significant during quiescence, and has not been included here.
  
   \begin{figure}
   \centering
   \includegraphics[angle=-90,width=\columnwidth]{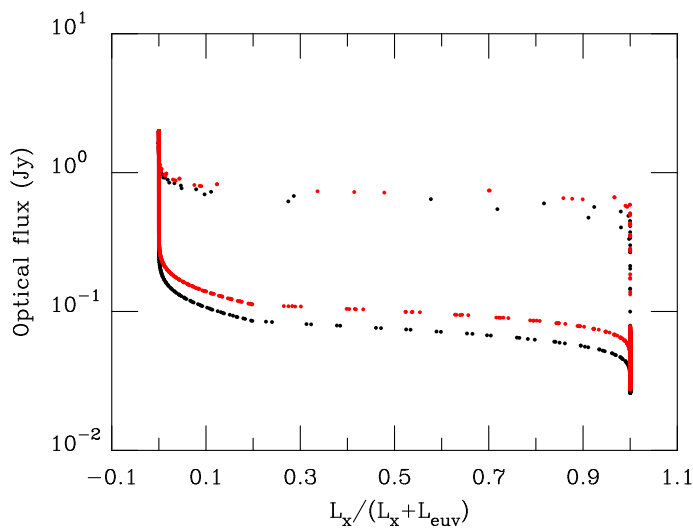}
   \caption{Hardness intensity diagram from the DIM. Several outbursts cycles are represented here, and each point is separated by approximately 3 hr. The unirradiated case is shown in black, the irradiated case in red.}
   \label{fig:hysteresis}%
   \end{figure}

Figure \ref{fig:optical} shows that the disc luminosity at time R during the rise phase is much larger than the luminosity at time D during decay; yet, the accretion rates at the inner disc edge are comparable in both cases. The reason for this is that during rise, the accretion rate at the inner edge of the disc increases on a viscous time, while the heating front propagates outwards on a much shorter thermal time. On the other hand, during the decay phase, the cooling front propagates inwards at a slower speed than a heating front \citep{mhs99}. For a given disc optical luminosity, and hence a given position of the transition front, the accretion rate should be smaller during rise than during the decay. It is therefore not surprising that an hysteresis such as the one seen by \citet{krkf08} is found in dwarf novae outbursts.

Figure \ref{fig:hysteresis} shows the hardness vs intensity diagram expected for a dwarf nova such as SS Cyg as predicted by the DIM. Here, the x-axis is the ratio between the X-ray and the X-ray-plus-EUV (defined as $\lambda$ smaller than 130 \AA) luminosities. Several outbursts are represented on this diagram, and it is worth noting that both the long and short outbursts follow the same track, despite the fact that the accreted mass is different in each case (10\% vs. 40\% of the total disc mass is accreted during short and long outbursts respectively). Our simulations have shown that short outbursts would produce different tracks on such a diagram only if the heating front were not able to reach the outer edge of the disc and the peak luminosity was much smaller than for long outbursts -- which is not the case for SS Cyg. We have found that this happens when e.g. the primary mass is large and the truncation radius is small. The irradiated and unirradiated cases are very similar; irradiation slightly reduces the amplitude of the hysteresis, because some optical flux, proportional to  $\dot{M}_{\rm acc}$ (and hence depending only on the ratio $L_{\rm X}/(L_{\rm X}+L_{\rm euv})$) is added. But because irradiation is negligible in the outer parts of the disc which contribute to most of the optical flux, this effect is small.

This figure can be directly compared with Fig. 1 from \citet{krkf08}. It is interesting to note that the general characteristics of our model-generated diagrams are rather similar to the observational diagram. The maximum optical luminosity is, as expected, close to the observed value. The transition X-ray / EUV on the upper branch occurs at about the right optical luminosity, which means that the critical $\dot{M}_{\rm acc}$ for the transition between optically thin /optically thick X-ray emission is not very different from the value used here (2 10$^{16}$ g/s). A slightly larger value, would have given an even better agreement. Note also that the transition is quite sharp. The optical luminosity on the lower branch is also quite comparable to the observed one; note, however, that the transition on the lower branch is smoother than that on the upper branch, but not quite as smooth as observed. This means that the exponential cut-off of the X-ray luminosity is probably too sharp. 

In order for this comparison to be meaningful, the spectral model has to be as realistic as possible. Modelling the contribution of the inner and outer disc and of the boundary layer is relatively simple, in contrast with the LMXB case (see below). It should be stressed that the hysteresis would appear in any case when considering the optical luminosity and any quantity linked to the mass accretion rate at the inner edge of the disc, and the conclusion that the DIM is responsible for the observed hysteresis is in fact very solid and do not depend on the detailed spectral modelling of the inner flow.

   \begin{figure}
   \centering
   \includegraphics[angle=-90,width=\columnwidth]{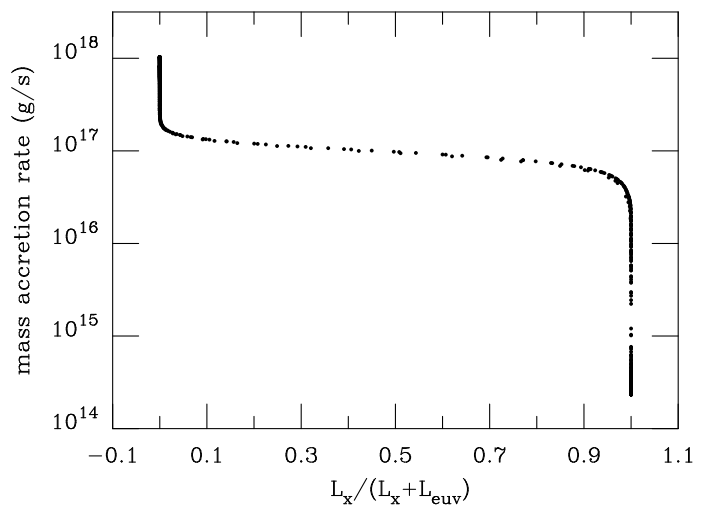}
   \caption{Hardness mass-accretion-rate onto the compact object diagram. As in Fig. \ref{fig:hysteresis}, several cycles are plotted on this diagram. Such a figure also represents the predictions of the DIM alone for low-mass X-ray binaries.}
   \label{fig:HID}%
   \end{figure}
   
Figure \ref{fig:HID} shows the same diagram as Fig. \ref{fig:hysteresis}, but we here plot the mass accretion rate onto the white dwarf instead of the optical luminosity. As can be seen, the hysteresis has disappeared and there is a univocal relation between $\dot{M}_{\rm acc}$ and the hardness ratio. This is merely due to the fact that the innermost part of the system which emits the X-rays and the EUV have very small characteristic time-scales and are therefore in a quasi steady state situation, with a local mass transfer rate determined by the outer parts of the disc which have a much longer viscous time. The precise shape of the dependence between $\dot{M}_{\rm acc}$ and the hardness ratio depends on the various assumptions made for the boundary layer and is therefore quite uncertain; but the very existence of the hysteresis in the hardness -- optical luminosity diagram is not affected by these assumptions. 
   
For the sake of completeness, we also calculated the hardness intensity diagram we obtain when we do not assume that the disc is truncated in quiescence, all other parameters being identical. The light-curves we find are relatively different to the truncated case: we obtain only one type of outbursts, similar to the short ones in Fig. \ref{fig:lightcurve}, separated by 47 days, and the peak optical luminosity is smaller. Yet the hardness intensity diagram we get in both cases are almost identical. This should not come as a surprise, since this diagram is largely determined by the viscous time in the outer parts of the disc which is responsible for the delay between the rise in the optical and the increase of the accretion rate onto the white dwarf.

\section{The case of LMXBs}

In LMXBs, the DIM provides a natural explanation for the occurrence of SXT outbursts, but the DIM focus is essentially on the outer regions of the disc, where the effective temperature is of order of a few thousands degrees, and where the instability occurs. Moreover, standard geometrically thin, optically thick Shakura-Sunyaev accretion discs are unable to produce the observed hard X-rays observed in LMXBs. The formation of an advection dominated flow when the mass transfer rate is low \citep[ADAF,][]{emn97} is an option which got strong support when realizing that the accretion disc had to be truncated for the DIM to be applicable to SXTs \citep{dhl01}. It remains, however, that there is no widely accepted model that can account for the spectral transitions of LMXBs, and henceforth for providing theoretical spectra of accretion discs for given physical parameters \citep[see e.g.][for recent alternative models]{smr13,ns14}. 

In addition, modelling the HID for LMXBs requires, in contrast with the CV case, an accurate modelling of the spectra of each of the components of the flow. This is because the energy bands used to determine the hardness ratio are adjacent whereas in the CV case, the EUV and X-ray bands are well separated and originate from well defined regions. And indeed, spectral models of accretion discs incorporating the effects of local dissipation in the vertical structure of the disc differ significantly from black bodies \citep{tb13}.

Despite these difficulties, it is, however, simple to show that the hysteresis in LMXBs is different from the one observed in CVs. The viscous time in the regions of the accretion disc where X-ray are produced is
\begin{equation}
t_{\rm visc} = \frac{r^2}{\nu} = \frac{1}{\alpha} \frac{v_{\rm K}^2}{c_{\rm s}^2} = \frac{1}{2\alpha} \frac{c^2}{c_{\rm s}^2} \frac{r_{\rm s}}{r}
\end{equation}
where $\nu$ is the viscosity, $v_{\rm K}$ is the Kepler viscosity and $r_{\rm s}$ is the Schwartzschild radius. For X-rays to be emitted, the effective temperature has to be larger than $10^7$ K, and hence the viscous time is less than:
\begin{equation}
t_{\rm visc} < 10^5 \frac{1}{\alpha} \frac{r_{\rm s}}{r} \; \rm s
\end{equation}
i.e. is always less than a day or a few days. On the transient time scale, the inner part of the disc will therefore be in a quasi steady-state situation, with a local mass transfer being imposed by the external parts of the disc which evolve on a much longer viscous time. This then requires that additional mechanisms are responsible for the existence of two different accretion flows for a given X-ray luminosity. Many such mechanisms have been proposed \citep[see e.g.][for recent examples]{ba14,ns14,c16}; they are clearly disconnected from the DIM.

\section{SS Cyg: an hysteresis in the $L_{\rm X} - L_{\rm EUV}$ domain?}

  \begin{figure}
   \centering
   \includegraphics[angle=0,width=\columnwidth]{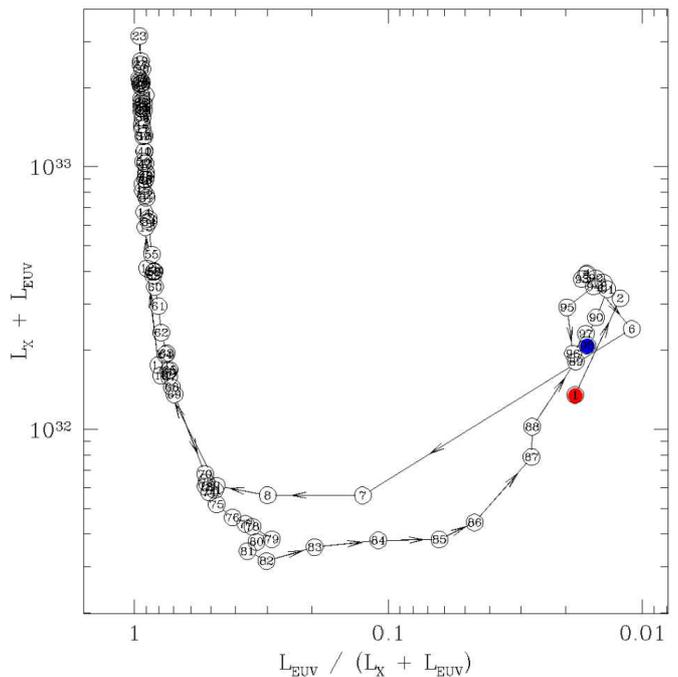}
   \caption{Observed HID for SS Cyg when considering the EUV/X-ray domain.}
   \label{fig:observations}%
   \end{figure}

The explanation by the DIM of the hysteresis observed by \citet{krkf08} does not of course mean that in CVs an hysteresis similar to the one observed in LMBXs does not exist, especially that for SS Cyg the white-dwarf jet-formation criterion by \citet{sl04} is not satisfied. But such a hysteresis should relate quantities  characterizing the state of the inner disc. This is, however, difficult to observe, as one needs to measure the accretion rate at the inner edge, while most of the related emission is in the UV - EUV band, thus heavily absorbed, and the bolometric corrections are important and uncertain.

In Fig. \ref{fig:observations} we show the HID obtained for SS Cyg when plotting the EUV luminosity versus the hardness defined as the ratio of the EUV and X-ray plus EUV luminosities, which should be representative of the accretion rate onto the white dwarf. The data are taken from \citet{wmm03}. Unfortunately the hysteresis hinges on 2 data-points but has a magnitude larger than 0.3 dex. A similar picture can be seen if one plots the X-rays as a function of hardness. If one combines both the X-rays and the EUV to get something like a bolometric luminosity, one obtains a hardness intensity diagram not dissimilar to that seen in XRBs. As we do not know the variable bolometric correction factors, it is unclear how these plots relate to a proper hardness intensity diagram for the inner accretion flow. But the data is indicative that such a hysteresis can exist. If it were confirmed, such an hysteresis would not be related to the DIM, as shown by Fig. \ref{fig:HID}, and would in fact bring no constrain on the instability responsible for dwarf nova outbursts; any mechanism accounting for the outburst of dwarf nova outburst \citep[such as e.g. the mass transfer instability model,][]{b73} would produce exactly the same HID because the inner disc is quasi-steady. This hysteresis would, however, be key in understanding the physics of accretion in the immediate vicinity of the white dwarf.

\section{Conclusions}

The DIM naturally produces an hysteresis in a diagram plotting the optical luminosity versus the X-ray / EUV hardness or any quantity which is directly related to the accretion rate onto the white dwarf. This hysteresis is simply due to the fact that it takes a viscous time for the mass transfer rate to vary at the inner disc edge whereas the optical luminosity varies on the much shorter thermal time. This is very different from the LMXB case, where the hysteresis relates soft and hard X-rays which are both emitted in the innermost parts of the disc, which, on the observational time-scales can then safely be assumed to be in a steady state situation, meaning that for a given local mass accretion rate, two quasi-stable solutions can exist. In addition, in CVs the hysteresis relates emission from various parts of the accretion disc whereas in case of LMXBs it might relate emission from the disc and from a hot extended flow not being part of the disc.

The X-ray / EUV hardness-intensity diagram for SS Cyg, which is similar to the classical HID for LMXBs shows clear indications that an hysteresis is present, which cannot be accounted for by the DIM, but is indicative of several components of the accretion flow being present for a given mass accretion rate in the inner disc, and whose existence depend on the history of the system. This hysteresis needs, however, to be confirmed.

It is difficult, but not impossible to test the models with current instrumentation. The major difficulty is accessing the EUV (CVs) and/or soft X-rays (XRBs). There are two problems there: (i) they are so easily absorbed away by the ISM; (ii) there are few observatories/instruments capable of observing in this regime. Overcoming the ISM issue -- which {\em especially} affects the EUV -- requires observing systems specifically selected to have low NH columns. Such systems are rare, but they do exist. Perhaps the best example of such a system is VW Hyi, which has a column of only $N_{\rm H} = 6 \; 10^{17}$ cm$^{-2}$ \citep{pmw90}. There are also others that may in principle be observable (e.g. SS Cyg was successfully observed with EUVE). Overcoming the instrumental issue is harder.  EUVE does not exist any more. Of the currently available space observatories/instruments, the only one which could be used for CVs in this regard is the Chandra/LETG combination. In fact, this has already been used to study WZ Sge \citep{m04a} and SS Cyg \citep{m04b}. GOALS (the Great Observatories Accretion Legacy Survey) is an observational campaign being developed, which is aimed at following an entire dwarf nova outburst from rise to decline across all wavelengths, covering X-rays, EUV, FUV, NUV, optical, NIR, radio. A GOALS-Pathfinder campaign with Chandra -- aimed at establishing whether two candidate targets (RX And and YZ Cnc) are detectable with the LETG -- is already approved and underway.

\begin{acknowledgements}
This work was supported by a National Science Centre, Poland grant 2015/19/B/ST9/01099. JPL was supported by a grant from the French Space Agency CNES.
\end{acknowledgements}

\end{document}